\documentclass[preprint,aps,prd,nofootinbib,superscriptaddress]{revtex4}
\usepackage[pdftex]{graphicx,color}
\usepackage{slashed}
\usepackage{amsmath,amssymb}
\usepackage{hhline}
\usepackage{ulem}
\usepackage{epstopdf}

\newcommand{\ba}{\begin{eqnarray}}
\newcommand{\ea}{\end{eqnarray}}
\newcommand{\be}{\begin{equation}}
\newcommand{\ee}{\end{equation}}

\newcommand{\Bbar}{\,\overline{\!B}{}}
\newcommand{\Dbar}{\,\overline{\!D}{}}
\newcommand{\Kbar}{\,\overline{\!K}{}}
\def\B0bar{\Bbar{}^0}
\def\D0bar{\Dbar{}^0}
\def\K0bar{\Kbar{}^0}
\def\DDbar{D^0\Dbar{}^0}
\usepackage[center]{subfigure}

\begin{document}
\begin{flushright}
SI-HEP-2017-12\\
QFET-2017-09\\
WSU-HEP-1709\\
\end{flushright}
\title{\boldmath Direct CP asymmetry in 
$D\to \pi^-\pi^+$ and $D\to K^-K^+$ in  QCD-based approach}

\author{\bf Alexander Khodjamirian}
\affiliation{Theoretische Physik 1, Naturwissenschaftlich-Technische 
Fakult\"at,\\
Universit\"at Siegen, D-57068 Siegen, Germany}

\author{\bf Alexey A.\ Petrov}
\affiliation{Theoretische Physik 1, Naturwissenschaftlich-Technische 
Fakult\"at,\\
Universit\"at Siegen, D-57068 Siegen, Germany}

\affiliation{Department of Physics and Astronomy\\
        Wayne State University, Detroit, MI 48201, USA}

\affiliation{Michigan Center for Theoretical Physics\\
        University of Michigan, Ann Arbor, MI 48196, USA}

\begin{abstract}
\noindent
We present the first QCD-based calculation of  hadronic matrix elements with penguin topology 
determining direct CP-violating asymmetries in $D^0\to \pi^-\pi^+$ and $D^0\to K^- K^+$ nonleptonic 
decays. The method is based on the QCD light-cone sum rules and does not rely on any model-inspired 
amplitude decomposition, instead leaning heavily on quark-hadron duality. We provide a Standard 
Model estimate of the direct CP-violating asymmetries in both pion and kaon modes and their difference 
and comment on further improvements of the presented computation.
\end{abstract}

\maketitle

\section{Introduction}\label{Sec:Introduction}

Despite years of intense experimental efforts, CP-violation 
has never been unambiguously observed in the decays of up-type quarks.
In the Standard Model (SM) this fact can be explained by the
suppression of all CP-violating amplitudes resulting from the
smallness of relevant Cabbibo-Kobayashi-Maskawa (CKM) matrix
elements. To make matters worst, accurate predictions of up-type CP-violating observables 
are hard to obtain, since the necessary hadronic matrix elements are dominated by 
long-distance contributions. In order to calculate these  matrix elements one needs to employ a 
QCD-based method that deals with nonperturbative effects in a model-independent manner. 
In this letter we shall calculate CP-violating observables in exclusive singly Cabibbo-suppressed (SCS) 
decays of $D$-mesons using a variant of light-cone QCD sum rules (LCSRs).

Observables that are sensitive to $CP$-violation are most often
written in terms of asymmetries \cite{BigiSandaBook},
\be
\label{CPVasym}
a_{\rm CP}(f) = \frac{\Gamma(D\to f)-\Gamma(\overline{D}\to\overline{f})}{
\Gamma(D\to f)+\Gamma(\overline{D}\to\overline{f})},
\ee
formed from the partial rates of a $D$-meson decay to a final state $f$ and of its CP-conjugated counterpart. 
Depending on the initial state, the asymmetry in Eq.~(\ref{CPVasym}) could be a function of time, 
if $\DDbar$-mixing is taken into account.
 
The measured time-integrated asymmetry contains a {\it direct} component (see, e.g., \cite{CPdirect}), which will 
be the main focus of this paper. Direct $CP$ asymmetry occurs when the absolute values of the $D\to f$ decay 
amplitude, which we denote by $A_f \equiv A(D\to f)$, and of the corresponding CP-conjugated amplitude 
${\overline A}_{\bar f} \equiv A(\overline D \to \overline f)$ are different. This can be realized 
if the decay amplitude $A_f$ can be separated into at least two different parts,
\be\label{DirectAmpl}
A_f =
A_f^{\rm (1)} e^{i \delta_1} e^{i \phi_1} +
A_f^{\rm (2)}  e^{i \delta_2} e^{i \phi_2},
\ee
where $\phi_1\neq \phi_2$ are the weak phases (odd under $CP$),
and $\delta_1\neq \delta_2$ are the strong phases (even under $CP$). 
The CP-violating asymmetry is then given by
\be\label{CPA1}
a_{\rm CP}^{\rm dir} (f) \propto  \frac{A_f^{\rm (1)}}{A_f^{\rm (2)}} \ \sin(\delta_1-\delta_2) \sin(\phi_1-\phi_2).
\ee
The amplitude pattern of Eq.~(\ref{DirectAmpl}) naturally emerges
in the SCS nonleptonic decays such as $D^0\to K^-K^+$ and $D^0 \to \pi^-\pi^+$.

A model-independent computation of the amplitude ratio and the strong phase difference in Eq.~(\ref{CPA1})
is a daunting task in charm physics (for reviews see e.g.,\cite{CPreview}). In SM, the sizes of direct CP-violating
asymmetries for any final state are always proportional to a small combination of the CKM factors 
$V_{cb}^* V_{ub}$, which ensures that $a_{\rm CP}^{\rm dir} (f)$ is small even if the maximal
strong phase difference is assumed. Hence, if a larger value of CP-violating asymmetry is observed, it would provide a 
``smoking gun'' signal of new physics in the charm sector of quark-flavor transitions.
However, due to a very uncertain hadronic input, the available estimates of $a_{\rm CP}^{\rm dir} (f)$ obtained 
with various degree of model dependence are mostly qualitative, predicting
this asymmetry in the ballpark of a per mille.

Direct and indirect components of $CP$-violating asymmetries in the decays of neutral $D$-mesons
can be separated with a careful time-dependent analysis. Since the indirect components to a 
large extent are independent of the final state, it is advantageous to consider {\it differences} of 
direct CP-violating asymmetries $\Delta a_{\rm CP}^{\rm dir}$, 
\be\label{DelAcp}
\Delta a_{\rm CP}^{\rm dir} = a_{\rm CP}^{\rm dir} (f_1) - a_{\rm CP}^{\rm dir} (f_2).
\ee
This difference is especially interesting if $CP$ asymmetries in the subtracted amplitudes are 
predicted to have opposite signs. This is in fact realized in SM for $f_1=K^-K^+$ and $f_2=\pi^-\pi^+$ final states.

Earlier experimental results seemed to indicate a somewhat large asymmetry (\ref{DelAcp}) for these final states,
with values reaching the order of $-1.0\%$. If confirmed, this would have indicated a possible new physics contribution 
to flavor-changing neutral currents (FCNC) in charm sector \cite{NP_predictions} or a previously unaccounted 
SM contributions \cite{SM_predictions}. 

Current measurements, however, yield significantly lower values, with an average \cite{PDG}
\be 
\Delta a^{dir}_{CP}= 
(-0.12 \pm 0.13)\%\,,
\label{eq:acpdel_PDG}
\ee
including the most accurate measurement to date by LHCb collaboration using
the $D^*$ tag \cite{Aaij:2016cfh},
\be 
\Delta a^{dir}_{CP}=(-0.10 \pm 0.08\pm 0.03)\%\, 
\label{eq:acpdel_LHCb}
\ee
in a qualitative agreement with the SM expectations \cite{NewSM_predictions}.

Combination of indirect and direct CP-asymmetries for 
$K^-K^+$ and $\pi^-\pi^+$ final states have also been separately measured. 
Averaged over several experiments they are reported to be \cite{Amhis:2016xyh}
\be
a_{CP}(K^-K^+)= (-0.16\pm 0.12) \%\,,~~
a_{CP}(\pi^-\pi^+)= (+0.00\pm 0.15)\%\,,
\label{eq:acppi_LHCb}
\ee
while the most recent combinations of measurements by LHCb collaboration read \cite{Aaij:2016dfb}
\begin{eqnarray}
a_{CP}(K^-K^+) &=& (0.04\pm 0.12 \ (\mbox{stat}) \pm  0.10 \ (\mbox{syst})) \%\,,
\nonumber \\
a_{CP}(\pi^-\pi^+) &=& (0.07\pm 0.14 \ (\mbox{stat}) \pm  0.11 \ (\mbox{syst}))\%\,.
\label{eq:acppi_LHCb2}
\end{eqnarray}
Both asymmetries imply a very small effect, consistent with zero within current experimental uncertainties. 
Note again that in the SM opposite signs are expected for the asymmetries in 
the $K^-K^+$ and $\pi^-\pi^+$  channels. 

New results with smaller experimental uncertainty are expected from the Run II LHCb data, as well as from 
the Belle II experiment. A model-independent calculation with controlled theoretical uncertainties 
of direct CP-violating asymmetries in SM is, therefore, compellingly needed. 
However, this task necessitates a calculation of hadronic matrix elements
with strongly interacting and energetic two-meson final states, which is a  big challenge 
even for the most advanced lattice QCD methods. In this situation
even an order-of-magnitude QCD-based estimate of the hadronic input should become useful
to reliably constrain the expected SM contribution to the CP-violating asymmetries.

The aim of this letter is to estimate the hadronic matrix elements relevant for the direct 
CP-violating asymmetries in $D^0\to P^-P^+$ decays ($P=\pi,K$),
employing a computational method which combines the light-cone operator-product 
expansion (OPE) in QCD and hadronic dispersion relations. 
More specifically, we use the approach developed in \cite{AK01} for
$B\to \pi\pi$ nonleptonic decays, in particular its application 
to the penguin-topology matrix elements \cite{KMM03}.   
We define matrix elements of ``penguin topology" as those of the weak effective 
4-quark operator containing a quark-antiquark pair not present in the valence-quark
content of the final $P^-P^+$ state.

In what follows, we identify the hadronic matrix elements with 
penguin topologies which are needed to estimate the direct $CP$ asymmetry in 
$D^0 \to \pi^- \pi^+$ and  $D^0 \to K^- K^+$ decays. 
We then briefly describe the method of Refs. \cite{AK01,KMM03}, adapting it to the 
$D^0 \to P^- P^+$ decays. The main result is the QCD LCSR for the
$D^0 \to P^- P^+$ hadronic matrix elements with the penguin topology. 
The calculation, which takes into account $SU(3)_{F}$-violating 
${\cal O}(m_s)$ effects, is valid at large invariant mass $P^2$ of the $\pi^-\pi^+$ and/or $K^-K^+$ final state. 
Applying quark-hadron duality, the result is analytically continued to the physical point $P^2=m_D^2$. 
The rest of this letter contains numerical analysis and our estimate of the direct CP-asymmetry in 
$D^0 \to P^- P^+$ decays, followed by a concluding discussion.

\section{$D^0\to \pi^-\pi^+$ and $D^0\to K^-K^+$ decay amplitudes}\label{Sec:ampl}

The singly Cabibbo suppressed decays of charmed mesons are driven by an effective Hamiltonian 
\begin{equation} \label{eq:Heff}
{\cal H}_{\rm eff}= \frac{G_{F}}{\sqrt{2}}
\left[
\sum_{q=d,s} \lambda_q \left(C_1 {\cal Q}_{1}^q + C_2 {\cal Q}_{2}^q\right) 
-\lambda_b\!\!\!\!\sum\limits_{i=3,...,6,8g} C_{i} {\cal Q}_{i}
\right] \,,
\end{equation}
where the products of CKM matrix elements $\lambda_q$ are defined as
\begin{equation}
\lambda_{q}=V_{uq}V^*_{cq}, \quad \mbox{with} \ \ q=d,s,b \,.
\label{eq:lam}
\end{equation}
Unitarity of the CKM matrix\footnote{Note that here we need to include $O(\lambda^5)$ terms in the 
Wolfenstein parameterization for the combinations $\lambda_{d,s}$ of CKM matrix elements (see e.g. \cite{CKM}).} 
implies that
\begin{equation}\label{eq:lamb}
\sum_{q=d,s,b} \lambda_q=0 \ \ \mbox {or}  \ \ \lambda_d=-(\lambda_s+\lambda_b).
\end{equation}
Since the goal of this paper is to capture the dominant contributions to the decay 
amplitudes in the SM, we shall only take into account the effective current-current operators,
\begin{eqnarray}\label{eq:012}
{\cal Q}_1^q =
\left(\bar{u}\Gamma_\mu q\right) 
\left(\bar{q}\Gamma^\mu c\right), \quad\quad
{\cal Q}_2^q = \left(\bar{q}\Gamma_\mu q\right) 
\left(\bar{u}\Gamma^\mu c\right)\,,
\end{eqnarray}
where $\Gamma_\mu = \gamma_\mu (1-\gamma_5)$, and $q=d,s$. 
We shall neglect the penguin operators ${\cal Q}_{i=3,...,6,8g}$ with small Wilson coefficients.

Furthermore, we introduce a compact notation for the linear combination of 
the operators (\ref{eq:012}) with their Wilson coefficients and the Fermi constant, 
\begin{equation} \label{eq:oper}
{\cal O}^{q}\equiv 
\frac{G_F}{\sqrt{2}}\sum\limits_{i=1,2} C_i {\cal Q}^q_i\,, \quad \ \mbox{with} \ \ q=d,s . 
\end{equation}
The dominant contribution to the two-body $D^0\to P^-P^+ $ nonleptonic decay amplitudes 
is given by the hadronic matrix elements of ${\cal O}^q$,
\begin{eqnarray}
A(D^0\to \pi^-\pi^+) &=&
\lambda_d\langle \pi^-\pi^+|{\cal O}^{d}|D^0\rangle +
\lambda_s\langle \pi^-\pi^+|{\cal O}^{s}|D^0\rangle \,, 
\label{eq:amplDpipi}
\\
A(D^0\to K^-K^+) &=&
\lambda_s\langle K^-K^+|{\cal O}^{s}|D^0\rangle +
\lambda_d\langle K^-K^+|{\cal O}^{d}|D^0\rangle \,. 
 \label{eq:amplDKK}
\end{eqnarray}
Applying the CKM unitarity relation of Eq.~(\ref{eq:lamb}) to Eqs.~(\ref{eq:amplDpipi}) and (\ref{eq:amplDKK}), 
and subsequently adding and subtracting a term $\lambda_b \ \langle \pi^-\pi^+|{\cal O}^{s}|D^0\rangle$ 
to right-hand side of Eq.~(\ref{eq:amplDpipi}), we arrange the decay amplitudes in the following form,
\begin{eqnarray}
A(D^0\to \pi^-\pi^+) &=& 
-\lambda_s {\cal A }_{\pi\pi}
\left[1 
+ \frac{\lambda_b}{\lambda_s} \left(1+ r_{\pi} \exp(i\delta_{\pi}) \right) \right], 
\nonumber  \\ 
A(D^0\to K^- K^+) &=& 
\phantom{-} \lambda_s {\cal A }_{KK}
\left[ 1 - \frac{\lambda_b}{\lambda_s}  r_{K} \exp(i\delta_{K})
\right],
\label{eq:decomp1}
\end{eqnarray}
where we introduce compact notations for the ratios
\begin{eqnarray}
r_{\pi}=\left|\frac{{\cal P}_{\pi\pi}^s}{{\cal A}_{\pi\pi}}
\right|\,, ~~~~
r_{K}=\left|\frac{{\cal P}_{KK}^d}{{\cal A}_{KK}}\right| 
\label{eq:rK}
\end{eqnarray}
of hadronic matrix elements
\begin{eqnarray}
{\cal P}_{\pi\pi}^s= \langle \pi^-\pi^+|{\cal 
  O}^{s}|D^0\rangle\,,~~~~
{\cal P}_{KK}^d= \langle K^-K^+|{\cal 
  O}^{d}|D^0\rangle\,,
\label{eq:O12}
\end{eqnarray}
and 
\begin{eqnarray}
{\cal A }_{\pi\pi} &=& \langle \pi^-\pi^+|{\cal O}^{d}|D^0\rangle-
\langle \pi^-\pi^+|{\cal O}^{s}|D^0\rangle\,,
\nonumber\\
{\cal A }_{KK} &=& \langle K^-K^+|{\cal O}^{s}|D^0\rangle-
\langle K^-K^+|{\cal O}^{d}|D^0\rangle\,,
\label{eq:calApiK}
\end{eqnarray}
and denote by $\delta_{\pi (K)}$ the difference between the strong phases 
of the amplitudes ${\cal P}_{\pi\pi(KK)}^{s(d)}$ and ${\cal A}_{\pi\pi(KK)}$. 
In what follows, we do not attempt to calculate the amplitudes ${\cal A}_{\pi\pi}$ 
and ${\cal A}_{KK}$, having in mind their complicated form in terms
of hadronic matrix elements with various quark topologies.
Instead, as follows from Eq.~(\ref{eq:decomp1}), these amplitudes 
can be extracted to a reasonable precision from the measured partial widths of 
$D^0\to \pi^-\pi^+$ and $D^0\to K^-K^+$ decays, neglecting small parts of the 
amplitudes proportional to $\lambda_b$.   

It is instructive to discuss the flavor $SU(3)_F\,$-symmetry limit of the decay amplitudes in Eq.~(\ref{eq:decomp1}). 
In this limit separate hadronic matrix elements with pions and kaons in the final state are equal: 
${\cal A }_{KK} ={\cal A }_{\pi\pi}$, $r_K=r_\pi$ and $\delta_K=\delta_\pi$. Still, with $\lambda_b\neq 0$,  
the decay amplitudes in Eq.~(\ref{eq:decomp1}) differ from each other by $O(\lambda_b)$ terms.
This difference can be easily understood in the $U$-spin symmetry limit, in which
the initial $D^0$-state is a $U$-singlet. The effective Hamiltonian of Eq.~(\ref{eq:Heff})
transforms as a combination of a $U$-triplet and $U$-singlet, the latter being proportional to $\lambda_b$, 
so that there are two $U$-spin invariant amplitudes contributing to  $A(D\to \pi^-\pi^+)$ and $A(D\to K^-K^+)$ 
with different coefficients\,\footnote{Alternatively,  one may use general $SU(3)_F$-expansion  
of these amplitudes (see e.g., \cite{Hiller:2012xm}) expressing them via two 
independent combinations of reduced matrix elements.}. 

The representation in Eq.~(\ref{eq:decomp1}) has the advantage that the 
parts of the amplitudes proportional to $\lambda_b/\lambda_s$ generate direct 
$CP$-violating asymmetries, due to the weak phase contained in this combination 
of CKM parameters. The asymmetries will vanish if either the ratio 
$r_{\pi,K}\to 0$ or the strong phase $\delta_{\pi,K} \to 0$. Hence, computation of the 
ratios $r_{\pi,K}$ and the phases $\delta_{\pi,K}$ will result in the prediction 
of direct CP-violating asymmetries of Eq.~(\ref{DelAcp}).

It is important to note that we will not be using the expansion of 
the decay amplitudes in flavor topology diagrams, which is frequently 
employed in the analysis of two-body 
nonleptonic decays of charmed meson, starting from the earlier papers \cite{quarkdiag}
(for a more recent analysis see, e.g, \cite{Rosner2010}).
In that approach, flavor symmetries and experimental data are used to 
fit the ``topological amplitudes". In our calculation, such expansion 
is unnecessary, first of all because we are only calculating the ``penguin topology'' 
matrix elements in Eq.~(\ref{eq:O12}), estimating the 
dominant part of the decay amplitudes from experimental measurements. 

Most importantly, in the commonly adopted convention, since the part of decay amplitude 
containing weak phase is suppressed by a very  small CKM coefficient $\lambda_b$,  
it is not feasible to extract this part from the experimentally observed decay rates, but rather 
calculate it directly, as it is done in this letter.

\section{Hadronic Matrix elements from LCSR\scriptsize{s}}

Here we adapt for nonleptonic $D$-decays the approach to compute hadronic amplitudes for 
$B\to \pi\pi$ decays suggested in \cite{AK01} and based on the method of QCD LCSRs \cite{lcsr}.
In particular, we will readily use the calculation of hadronic matrix elements of 
$B\to \pi\pi$ decays with charm penguin topology performed in \cite{KMM03}.
We start from the $D^0\to \pi^-\pi^+$ transition aiming at estimating the $s$-quark ``penguin" 
contribution ${\cal P}_{\pi\pi}^{s}$ to this decay amplitude. Following \cite{KMM03}, the starting 
object is the correlation function
\begin{eqnarray}
F_\alpha(p,q,k)=i^2\int d^4x e^{-i(p-q)x}\int d^4y  e^{i(p-k)y}
\langle 0 | \ T \left\{ j_{\alpha 5}^{(\pi)}(y) {\cal Q}_1^{s}(0) j_{5}^{(D)}(x) \right \}|\pi^+(q)\rangle
\nonumber\\
=(p-k)_\alpha F((p-k)^2,(p-q)^2,P^2) +
\dots,
\label{eq:corr}
\end{eqnarray}
where $j_{\alpha 5}^{(\pi)}=\bar{d}\gamma_\alpha\gamma_5 u$ and 
 $j_{5}^{(D)}=im_c\bar{c}\gamma_5 u$ are, respectively, the pion and $D$-meson
interpolating currents, sandwiched together with the four-quark operator 
${\cal Q}_1^{s}$ between the on-shell pion and vacuum state. 
The ellipsis denote the kinematical structures we do not use. Note that 
performing a Fierz transformation of the operator ${\cal Q}^s_1$, the 
combination of current-current operators entering the effective
Hamiltonian of Eq.~(\ref{eq:Heff}) is transformed:
\begin{eqnarray}
C_1{\cal Q}_1^s + C_2{\cal Q}^s_2= 2 C_1\widetilde{{\cal Q}}_2^s+ 
\left(\frac{C_1}{3} + C_2\right){\cal Q}^s_2\,,
\label{eq:Fiertz}
\end{eqnarray}
so that the color-octet operator 
\begin{eqnarray}
\widetilde{{\cal Q}}_2^s= \left(\bar{s}\Gamma_\mu \frac{\lambda^a}{2}s\right) 
\left(\bar{u}\Gamma^\mu\frac{\lambda^a}{2} c\right)\,,
\label{eq:coloroct}
\end{eqnarray}
is the only one contributing to the penguin amplitude in the adopted approximation.
Hence, we hereafter replace ${\cal Q}_1\to \widetilde{{\cal Q}}_2^s$  
in the correlation function. The resulting hadronic matrix element obtained from LCSR
below is related to the penguin matrix element:
\begin{equation}
{\cal P}^s_{\pi\pi} = \frac{2 G_F}{\sqrt{2}} \ C_1
\langle \pi^+\pi^-|  \widetilde{{\cal Q}}_2^s|D^0\rangle\,.
\label{eq:peng_norm}
\end{equation}
Following \cite{AK01} we introduce an auxiliary 4-momentum $k$ flowing from the vertex of the weak interaction  
in the correlation function (\ref{eq:corr}) and assume $k^2=0$, and, for simplicity,
$p^2=0$. We adopt massless $u$ and $d$ quarks and 
a massless pion ($q^2=0$) approximation, so that the invariant amplitude determining the 
correlation function depends on three invariant variables $(p-k)^2$, $(p-q)^2$ and $P^2=(p-q-k)^2$. 
In the spacelike region $|(p-k)^2|,|(p-q)^2|\gg \Lambda_{QCD}^2$, $|P^2 |\gg \Lambda_{QCD}^2$
the light-cone OPE expressed in terms of pion distribution amplitudes (DAs) is used 
to calculate the invariant amplitude $F((p-k)^2,(p-q)^2,P^2)$. 

\begin{figure}[t]\center
\includegraphics[scale=0.85]{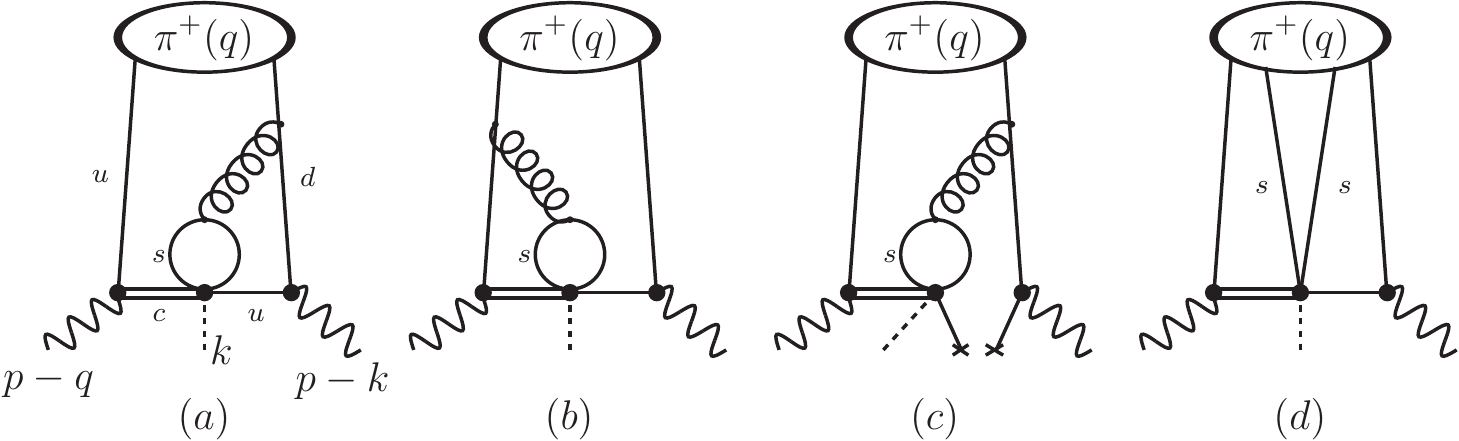} 
\caption{Diagrams contributing to the correlation function (\ref{eq:corr}).} 
\label{fig:1}
\end{figure}

As in \cite{KMM03}, the essential OPE diagrams are  the ones shown in 
Fig.~\ref{fig:1}. We remind the reader that these diagrams stem from 
the light-cone OPE of the correlation function. 
Each diagram contains a 
coefficient function (hard-scattering amplitude) calculated perturbatively and
convoluted with the pion distribution amplitudes (DAs) of growing twist and multiplicity. 
Therefore,  only the highly-virtual quark and gluon lines 
corresponding to near-light-cone separation (large spacelike momentum transfer) 
between the two external currents with momenta $p-q$ and $p-k$ are
included explicitly.
In particular, the $s$-quarks in the loops also have large virtualities.  
The small-virtuality quarks and gluons are by default included in the pion DAs.
Hence, for example, the diagrams with gluons emitted from the s-quark loop and absorbed in 
the pion DA should not be included in OPE~\footnote{ Similar diagrams with heavy 
$c$-quark loop in $B$ decay LCSRs as discussed in Ref.~\cite{KMM03} remain the part of OPE, 
albeit being the part of higher-twist power 
corrections.}. 

Here, as in \cite{KMM03}, we only keep the contributions of
two-particle twist-2 and twist-3  DAs 
and neglect all contributions of multiparticle pion DAs, 
since they correspond to higher-twist contributions to OPE and are suppressed 
by powers of characteristic large scales with respect to the lowest-twist 
contributions. In particular, we neglect contributions of 
diagrams where low virtuality (``soft") gluons are emitted 
from the virtual quarks and form quark-antiquark-gluon DAs of twist 
3 and 4. On the other hand, following \cite{KMM03} we take into account  
the diagram (c) containing the factorizable four-quark component of the pion DA, 
expressed via two-particle DA and vacuum quark condensate density.  
In the case of heavy charm-quark loop in  $B\to\pi\pi$-decays considered
in \cite{KMM03}  there is a second diagram of this
type with the $\bar{d}d$-condensate. In our correlation function such a diagram
would only have small-virtuality $s$-quarks  which cannot be resolved from 
the four-quark DAs, hence, it is absent in the context of OPE
(see also discussion in \cite{KMM03}).  
Note that the s-quark pair in the correlation function, 
being generated from  the V-A current, cannot itself form the $\bar{s}s$ condensate.
Hence, the small virtuality 
s-quark pairs can only form ``genuine" nonfactorizable four-particle pion DAs
originating from the nonlocal matrix element 
$\langle 0 | \bar{d}(y)\bar{s}(0)s(0)u(x)|\pi^+\rangle$ and described
by the diagram (d).
As mentioned above, we neglect these contributions.
%

The expressions for the diagrams in Figs.~\ref{fig:1}(a)-(c) 
are then taken from \cite{KMM03}, replacing $b$-quark by $c$-quark and $c$-quark 
in the loop by $s$-quark. 
Derivation of LCSR follows then the same three-step procedure as the one 
developed and discussed in detail in \cite{AK01}
and used  in \cite{KMM03}.
The first step is to use the hadronic dispersion relation for the invariant amplitude
$F((p-k)^2,(p-q)^2,P^2)$ in the variable  $(p-k)^2=s$ of the pion channel, 
keeping the variables $(p-q)^2$ and $P^2$ fixed and spacelike, applying 
quark-hadron duality approximation in this channel with the 
effective threshold $s_0^\pi$ and performing the Borel transformation 
$(p-k)^2\to M_1^2$. The second step is the 
transition from spacelike to large timelike $P^2=m_D^2$, 
assuming local quark-hadron duality. At this stage we obtain the relation: 
\begin{eqnarray}
i\int d^4x e^{-i(p-q)x}
\langle \pi^-(-q)\pi^+(p-k) |\ T\left\{ {\cal Q}_1^{s}(0) j_{5}^{(D)}(x)\right\}|0\rangle
\nonumber\\
=\frac{-i}{\pi f_\pi}\int\limits_0^{s_0^\pi}dse^{-s/M_1^2}\mbox{Im}_{s}F(s,(p-q)^2,m_D^2)\,,
\label{eq:disp1}
\end{eqnarray}
for the matrix element with  two-pion final state, produced by 
the product of the operator ${\cal Q}_1^{s}$ and $D$-meson interpolating
current  with fixed $(p-q)^2<0$; $f_\pi$ in the above is the pion decay constant.   
The final third step is to apply to the above matrix element 
the  hadronic dispersion relation in the variable $(p-q)^2=s'$, 
combined with the quark-hadron duality approximation in the $D$-meson 
channel with the effective threshold  $s_0^D$ and followed by the  
Borel transformation  $(p-q)^2\to M_2^2$. After that the auxiliary 
4-momentum $k$ vanishes for the $D$-meson pole term in the dispersion relation, leading
to the sum rule for the  $D^0\to \pi^-\pi^+$ hadronic matrix element: 
\begin{eqnarray}
\langle \!\pi^-\!(-q)\pi^+(p) |{\cal Q}_1^{s}|D^0(p-q)\rangle
\!=\!
\frac{-i}{\pi^2 f_\pi f_D m_D^2}\!\!\int\limits_0^{s_0^\pi}\!\!ds\,\!e^{-s/M_1^2}
\!\!\!\int\limits_{m_c^2}^{s_0^D}\! \!ds'\,\!e^{(m_D^2-s')/M_2^2}
\mbox{Im}_{s'}\mbox{Im}_{s}F(s,s'\!,\!m_D^2),
\label{eq:disp2}
\end{eqnarray}
where $f_D$ is the $D$-meson decay constant. The right-hand side of this expression is obtained by
computing the double imaginary part of the sum of OPE diagrams, resulting in the final form of LCSR,
\begin{eqnarray}
\langle \pi^-\pi^+|  \widetilde{{\cal Q}}_2^s|D^0\rangle
&=& i\frac{\alpha_s C_Fm_c^2}{8\pi^3m_D^2f_D}
\Bigg[\int\limits_0^{s_0^\pi}ds e^{-s/M_1^2}\int\limits_{u_0^D}^{1} 
\frac{du}{u}e^{\left(m_D^2-\frac{m_c^2}{u}\right)/M_2^2}
\nonumber\\
&\times&
\Bigg\{P^2\!\!\int\limits_0^1 dz I(zuP^2,m_s^2)\Bigg( z(1-z) 
\varphi_\pi(u)
\nonumber\\
&+&(1-z)\frac{\mu_\pi}{2m_c}
\Big[\left(2z+\frac{m_c^2}{uP^2}\right)u\phi^p_{3\pi}(u)
+\frac{1}{3}\left(2z-\frac{m_c^2}{uP^2}\right)
\left(\phi^\sigma_{3\pi}(u)-\frac{u\phi^{\sigma'}_{3\pi}(u)}{2}\right)
\Big]\Bigg)
\nonumber\\
&-&\frac{\mu_\pi m_c}{4}
\int\limits_0^1 dz I(-z\bar{u}m_c^2/u,m_s^2)\frac{\bar{u}^2}{u}\Big[
\left(1+\frac{3m_c^2}{uP^2}\right)\phi^p_{3\pi}(1)
+\left(1-\frac{5m_c^2}{uP^2}\right)\frac{\phi^{\sigma\prime}_{3\pi}(1)}{6}\Big]\Bigg\}
\nonumber\\
&+&
\frac{2\pi^2}{3} m_c(-\langle\bar{q}q\rangle)
\int\limits_{u_0^D}^{1} 
\frac{du}{u^2}e^{\left(m_D^2-\frac{m_c^2}{u}\right)/M_2^2}
\Bigg\{ I(uP^2,m_s^2)\Bigg(2\varphi_\pi(u)+\frac{\mu_\pi}{m_c}\Big[
3u\phi^p_{3\pi}(u) 
\nonumber
\\
&+&\frac{\phi^\sigma_{3\pi}(u)}{3}-\frac{u\phi^{\sigma \prime}_{3\pi}(u)}{6}\Big]\Bigg)
\Bigg\}\Bigg]_{P^2\to m_D^2}\,,
\label{eq:sr}
\end{eqnarray}
where $u_0^D=m_c^2/s_0^D$.  
The above expression is valid at large positive 
$P^2\gg s_0^\pi$, and the power corrections $O(s_0^\pi/P^2)$ are neglected. 
In Eq.~(\ref{eq:sr}) the quark-loop function is defined \cite{KMM03} as 
\be
I(\ell^2,m_q^2)=
\frac{1}{6} + \int\limits_0^1 dx x (1-x)
\ln\left [\frac{m_q^2- x(1-x)\ell^2}{\mu^2}\right]\,.
\label{eq:loop}
\ee
Note that the contributions from diagrams in Fig.~\ref{fig:1}a 
(Fig.~\ref{fig:1}b) are presented in the second and third (fourth) lines of 
Eq.~(\ref{eq:sr}), and the one from Fig.~\ref{fig:1}c occupies 
the fifth and sixth lines of Eq.~(\ref{eq:sr}).
The pion DAs entering the LCSR are $\varphi_\pi(u)$ and 
$\varphi_{3p}^\pi(u),\varphi_{3p}^\pi(u)$, of twist-2 and 
twist-3, respectively. They depend, as usual, on the fraction
of the longitudinal momentum of the pion $u$ ($\bar{u}=1-u$) 
carried by the quark (antiquark). 
We define $\phi^{\sigma \prime}_{3\pi}(u)=d \phi^{\sigma \prime}_{3\pi}(u)/du$
and adopt the truncated expansion in Gegenbauer polynomials 
$C_n^{(\alpha)}(u-\bar{u})$ for these DAs. We take
\ba
\varphi_\pi(u) = 6u\bar{u}\left(1+a_2^\pi C^{3/2}_2(u-\bar{u})+
a_4^\pi C^{3/2}_4(u-\bar{u}) \right)\,,
\label{eq:tw2DA}
\ea
for the twist-2 DA, and
\ba
\phi^p_{3\pi}(u) &=&1+30\frac{f_{3\pi}}{\mu_\pi f_\pi} C_2^{1/2}(u-\bar u)-
3\frac{f_{3\pi}\omega_{3\pi}}{\mu_\pi f_\pi} C_4^{1/2}(u-\bar{u}),
\nonumber
\\
\phi^\sigma_{3\pi}(u) &=& 6u(1-u)\left(1+5\frac{f_{3\pi}}{\mu_\pi f_\pi}
\left(1-\frac{\omega_{3\pi}}{10}\right)C_2^{3/2}(u-\bar u)\right)\,,
\label{eq:tw3DAs}
\ea
for the twist-3 DAs. Note that the Gegenbauer moments $a^\pi_{2,4}$ entering the twist-2 DA,
the normalization parameter $\mu_\pi=m_\pi^2/(m_u+m_d)$,  
and the parameters  $f_{3\pi}, \omega_{3\pi}$ of the non-asymptotic parts  
in twist-3 DAs  represent nonperturbative inputs determined from various sources. 
Their scale dependence is not shown for brevity and taken into account at leading order   
(see e.g., \cite{KMMO} for more details). In Eq.~(\ref{eq:sr}),
$\langle \bar{q}q\rangle\equiv \langle 0| \bar{u}u| 0 \rangle$  is the quark-condensate density.
Following other applications of QCD sum  rules, we use the $\overline{MS}$ values for 
quark masses $m_c$ and $m_s$ in the numerical analysis.

Similar considerations hold for the calculation of $D$-decay to the kaon final state.
For the kaon amplitude the analogous sum rule is obtained by replacing the following quantities
in the correlation function of Eq.~(\ref{eq:corr}): $\bar{d}\to \bar{s}$ in the interpolating current
(so that $j_{\alpha_5}^{(\pi)}\to j_{\alpha_5}^{(K)}$), the operator 
${\cal Q}_1^s \to {\cal Q}_1^d$, and the final state  $|\pi^+\rangle \to| K^+ \rangle$.
Correspondingly, the diagrams in Fig.~\ref{fig:1}  will change their flavor content, in particular,
the $s$-quark loop will be replaced by the $d$-quark loop, which is easily taken
into account by putting the mass of the internal quark in Eq.~(\ref{eq:loop}) to 
zero. In the rest of the diagrams we intend to include $O(m_s)$ 
and, correspondingly, $O(m_K^2)$ terms in order to 
assess the flavor $SU(3)_{F}$ violation in the hadronic matrix 
elements\,\footnote{ Certain parametrically smaller $O(m_s\Lambda_{QCD}/m_c^2)$ terms 
originating from the traces of the diagrams, cannot be captured in our analysis 
and demand recalculation of the whole OPE diagrams, 
a task for the future.}. 

Summarizing, the LCSR for the hadronic matrix element 
$\langle K^+K^-|  \widetilde{{\cal O}}_2^d|D^0\rangle$ is then
obtained from Eq.~(\ref{eq:sr}) by the following replacements,
\begin{eqnarray} \label{eq:srKK}
s_0^\pi &\to& s_0^K\,,~~ e^{-s/M_1^2} \to e^{(m_K^2-s)/M_1^2}\,, 
\nonumber
\\
e^{\left(m_D^2-\frac{m_c^2}{u}\right)/M_2^2} &\to& 
e^{\left(m_D^2-\frac{m_c^2+m_K^2\bar{u}u}{u}\right)/M_2^2}, ~~
I(..,m_s^2) \to I(...,0)\,,  ~~ 
\\
\varphi_\pi(u) &\to& \varphi_K(u), ~~
 ~~
\varphi^\pi_p(u) \to \varphi^K_p(u)\,,~~
\varphi^\pi_\sigma(u)\to\varphi^K_\sigma(u)\, ,
\nonumber
\end{eqnarray}
as well as by replacing $\langle \bar{q}q\rangle\to \langle\bar{s}s \rangle$,
 $\mu_\pi\to \mu_K$
and $f_\pi \to f_K$. In the interest of succinctness  
we only quote the kaon  DA of twist-2,
\be
\varphi_K(u)=6u\bar{u} \left(1+a_1^KC^{3/2}_1(u-\bar{u})+
a_2^KC^{3/2}_2(u-\bar{u}) \right)\,,
\label{eq:kaonDA}
\ee
where the Gegenbauer moment $a_1^K$ reflects the $SU(3)_F$-violating asymmetry 
of the $\bar{s}$ and $u$-quark average momentum fractions in the DA.
The expressions for kaon twist-3 DAs  can be found, e.g., in \cite{KMMO,BBL}. Apart from  the 
parameters $\mu_K$, $f_{3K}$, $\omega_{3K}$, analogous to the pion ones in
Eq.~(\ref{eq:tw3DAs}), these DAs also contain certain $O(m_s)$ corrections 
and an additional  $SU(3)_F$-asymmetry parameter $\lambda_{3K}$.

Note that here, similarly to the LCSR analysis of $B\to 2 \pi$ decays \cite{KMM03},
we neglect the penguin-annihilation contribution, which is expected to be 
$\alpha_s$- and power-suppressed. In principle, this contribution can be separately estimated 
using the same approach. That evaluation is however technically more involved, as it contains 
multiloop contributions.

In conclusion of this section we emphasize that the pion and kaon DAs of the lowest twist, 
as well as the interpolating currents of pion, kaon and $D$-meson all have a 
valence quark content of the corresponding hadrons, whereas the hadronic matrix elements 
${{\cal P}_{\pi\pi(KK)}^{s(d)}}$ are obtained from the diagrams where the quark pair in the 
relevant operator has a flavor different from the valence content. Hence, using the quark-hadron 
duality ansatz employed in the LCSRs, a hadronic matrix element with penguin topology is 
unambiguously identified, being ``protected'' at the level of correlation
function from additional quark-antiquark insertions. Indeed, in the OPE such insertions in DAs or in 
interpolating currents would  produce $\alpha_s$-suppressed and/or higher-twist (power suppressed)
contributions.

\section{Numerical results}

In order to estimate the size of the computed hadronic matrix elements we need to 
provide numerical inputs for various parameters used in this calculation. 
We make conventional choices for the renormalization and factorization 
scale $\mu$ in LCSRs, adopting $\mu= 1.5\pm 0.5$ GeV.
The intervals of the (universal) Borel parameter $M_1^2=1.0\pm 0.5 $ GeV$^2$ 
in the $\pi,K$ -meson channels, and the 
corresponding thresholds $s_0^\pi=0.7\pm 0.1 $ GeV$^2$, 
$s_0^K=1.2\pm 0.1 $ GeV$^2$  are inspired by the analysis of LCSRs 
for the pion and kaon electromagnetic form factors \cite{BKM}.
The corresponding parameters for $D$ meson channel,
$M_2^2=4.5\pm 1.0 $ GeV$^2$, $s_0^D=7.0\pm 0.5 $ GeV$^2$
are taken following the LCSR analysis of  $D\to \pi,K$ form factors \cite{KMMO}.
We display our choices for the remaining input parameters in Table~\ref{tab:input}.
They include the strong coupling, quark masses, quark-condensate densities and
the parameters of pion and kaon DAs, all rescaled to the adopted scale.
Finally, we use the value of 
$f_D=201\pm 13 $ MeV for the $D$-meson decay 
constant obtained from the 2-point QCD sum rule analysis in
\cite{Gelhausen:2013wia}, and the values 
$f_\pi = 130.5$ MeV and $f_K = 155.6$ MeV 
respectively \cite{PDG} for the pion and kaon decay constants. 
%
\begin{table}[t]
\begin{center}
\begin{tabular}{|c|c|}
\hline\hline
$~~~~~~~$ Parameter values $~~~~~~~$ & $~~~~~~~$Parameter rescaled  $~~~~~~~$ \\
 and references &  to $\mu=1.5$ GeV  \\
\hline\hline
$\alpha_s(m_Z)=0.1181\pm 0.0011$ ~\cite{PDG} & $ 0.351 $  \\
$\bar{m}_c(\bar{m}_c)=1.27 \pm 0.03 $ GeV \cite{PDG}&1.19  GeV \\
$\bar{m}_s(2\,\mbox{GeV})=96^{+8}_{ -4}$\,MeV  ~\cite{PDG} &105~MeV \\
$\langle \bar{q}q \rangle(2\,\mbox{GeV})=(-276 ^{+12}_{ -10}\,\mbox{MeV})^3$ 
\cite{PDG} &$ (-268\,\mbox{MeV})^3$ \\
$\langle \bar{s}s \rangle = (0.8\pm 0.3) \langle \bar{q}q \rangle $ ~\cite{Ioffe:2002ee}~~
&$(-249~\mbox{MeV})^3$ \\
\hline
$a_2^\pi(1\,\mbox{GeV})= 0.17 \pm 0.08$ ~\cite{Khodjamirian:2011ub}& 0.14  \\
$a_4^\pi(1\,\mbox{GeV})= 0.06 \pm 0.10$ ~\cite{Khodjamirian:2011ub}&0.045 \\
$\mu_\pi(2\,\mbox{GeV})= 2.48\pm 0.30$\,GeV ~\cite{PDG} &2.26\,GeV \\
$~~~$ $f_{3\pi}(1\,\mbox{GeV})=0.0045 \pm 0.015 $ GeV$^2$ ~\cite{BBL}$~~~$ & 0.0036\,GeV$^2$ \\
$\omega_{3\pi}(1\,\mbox{GeV})=-1.5 \pm 0.7 $ ~\cite{BBL}&-1.1 \\
\hline 
$a_1^K(1\,\mbox{GeV})=0.10 \pm 0.04$ ~\cite{Chetyrkin:2007vm}& 0.09  \\
$a_2^K(1\,\mbox{GeV})=0.25 \pm 0.15$ ~\cite{BBL} & 0.21 \\
$\mu_K(2\,\mbox{GeV})= 2.47 ^{+0.19}_{ -0.10}  $ GeV ~\cite{PDG} & 2.25  \\
$f_{3K}= f_{3\pi} $  &0.0036\,GeV$^2$  \\
$\omega_{3K} (1\,\mbox{GeV})=-1.2 \pm 0.7 $~\cite{BBL} &-0.99 \\
$\lambda_{3K} (1\,\mbox{GeV})=1.6 \pm 0.4 $ ~\cite{BBL}& 1.5 \\
\hline
\hline
\end{tabular}
\end{center}
\caption{ The input parameters used in the numerical analysis of LCSRs
and their values  at the renormalization scale $\mu=1.5$ GeV.}
\label{tab:input}
 \end{table}
%

With the chosen input, the results for the hadronic 
matrix elements calculated from the sum rule in Eq.~(\ref{eq:sr}) 
and from its analogue for the kaon channel are
\begin{eqnarray}
\langle ~\pi^+\pi^-| \widetilde{\cal Q}_2^s|D^0\rangle
&=& (9.50\pm 1.13 )\times 10^{-3} \exp[i(-97.5^{\scriptstyle o}\pm 11.6)]\, \mbox{GeV}^3 \, ,
\nonumber\\
\langle K^+K^-| \widetilde{\cal Q}_2^d|D^0\rangle
&=& (13.9\pm 2.70 )\times 10^{-3}\exp[i(-71.6^{\scriptstyle o}\pm 29.5)]\, \mbox{GeV}^3 \, ,
\label{eq:hme_num}
\end{eqnarray}
 where the imaginary parts generated by the quark loops should, 
in the quark-hadron duality approximation, reproduce the strong phases of these 
amplitudes\,\footnote{This is similar to the quark-loop generation of a strong phase 
in the heavy quark  decays \cite{Bander:1979px} and in particular, in the 
QCD factorization approach \cite{Beneke:2001ev}.}. 
 
Using Eq.~(\ref{eq:peng_norm}), we employ the value for the Wilson coefficient 
$C_1(\mu=1.5 \mbox{ GeV})= 1.25$ calculated at the same characteristic scale as the one 
used in LCSR for the hadronic matrix element. Finally, we obtain the estimate of the 
dimensionless penguin amplitudes,
\ba
|{\cal P}_{\pi\pi}^{s}|= (1.96 \pm 0.23) \times 10^{-7}\,,
\label{eq:rPres_pi}
\\
|{\cal P}_{KK}^{d}|= (2.86 \pm 0.56) \times 10^{-7}\,,
\label{eq:rPres_K}
\ea
for the pion and kaon final states, respectively. The uncertainties in 
Eqs.~(\ref{eq:hme_num}), (\ref{eq:rPres_pi}), and (\ref{eq:rPres_K}) 
are obtained by randomly varying input parameters (given above and in 
Table~\ref{tab:input}) within their adopted ranges interpreted as $1\sigma$-intervals. 
To this end,  a statistics of $10^4$  parameter combinations was generated
for each LCSR. We assume no correlation between various inputs which certainly makes the 
uncertainty estimate more conservative.
One has to emphasize that only parametrical uncertainties are taken into account here.
The approximation for the OPE diagrams we used  from 
\cite{KMM03} neglects small terms of $O(s^{\pi}_0/m_B^2)\sim 4\%$, 
hence additional corrections to the LCSRs at the level 
of $O(s^{\pi,K}_0/m_D^2)\sim 30\%$ cannot be excluded in the case of $D$-meson 
hadronic matrix elements. A dedicated calculation of the OPE diagrams 
is needed to include these corrections.
       
The values in Eqs.~(\ref{eq:rPres_pi}) and (\ref{eq:rPres_K})
will be used in the next section to estimate the direct CP-violating asymmetries 
and their difference in kaon and pion channels.

\section{Direct CP-violating asymmetry}\label{NumericalCP}

Neglecting $\DDbar$ mixing, the partial rates 
\begin{equation}
\Gamma(f) \equiv \Gamma(D^0\to P^-P^+)
\end{equation} 
for $P=\pi,K$ can be written as
\begin{equation}   
\Gamma(D^0\to P^-P^+)= \frac{p^*_P}{8\pi m_D^2}|A(D^0\to P^-P^+)|^2\,,
\label{eq:width}
\end{equation}
where $p^*_P$ is the decay 3-momentum in the $D$-meson rest frame.
In terms of the amplitude parametrization in Eq.~(\ref{eq:decomp1}), 
the direct part of the CP-asymmetry for the decay of a D-meson 
into kaons is
\begin{equation} \label{eq:acp1}
a_{CP}^{dir}(K^-K^+) = -\frac{2r_{b} r_{K} 
\sin \delta_{K} \sin \gamma}{1 - 2r_{b}r_{K}\cos\gamma \cos\delta_{K} +
r_{b}^2r_{K}^2}\,,
\end{equation}
while the same asymmetry for the decay of a $D$-meson into pions is
\begin{equation} \label{eq:acp2}
a_{CP}^{dir}(\pi^-\pi^+) = \frac{2r_{b} r_{\pi} 
\sin \delta_{\pi} \sin \gamma}{1+2r_{b}\cos\gamma (1+r_{\pi}\cos\delta_\pi) + 
r_{b}^2(1+2r_{\pi}\cos\delta_{\pi}+r_{\pi}^2)}\,.
\end{equation}
In Eqs.~(\ref{eq:acp1}) and (\ref{eq:acp2}) it was convenient to represent the ratio of 
CKM parameters as 
\begin{equation}
 \frac{\lambda_b}{\lambda_s}\equiv r_be^{-i\gamma},~~
\label{eq:lb}
\end{equation}
where, in terms of Wolfenstein parameters, the modulus is
\begin{equation}
r_b \equiv \left|\frac{V_{ub}V_{cb}^*}{V_{us}V_{cs}^*}\right|
= A^2\lambda^4\sqrt{\rho^2+\eta^2}+O(\lambda^6)\,.
\label{eq:WP}
\end{equation}
while the phase $\gamma=\arg(\rho+i\eta)$ coincides with the angle $\gamma$ of the 
unitarity triangle.

Due to the interplay of CKM coefficients in SM, the individual direct $CP$-violating 
asymmetries $a_{CP}^{dir}(K^-K^+)$ and  $a_{CP}^{dir}(\pi^-\pi^+)$ have opposite signs, as expected.
Taking the difference of Eqs.~(\ref{eq:acp1}) and (\ref{eq:acp2}), expanding the 
result in $r_b$, and keeping only the linear piece we obtain
\begin{eqnarray}
\Delta a_{CP}^{dir}= a_{CP}^{dir}(K^-K^+)-a_{CP}^{dir}(\pi^-\pi^+)= 
-2r_b \sin\gamma(r_K \sin\delta_K+r_\pi \sin \delta_\pi) + {\cal O} (r_b^2)\,.
\label{eq:deltaAcp}
\end{eqnarray}
In order to perform a numerical analysis of the above formula we take 
the central values of the Wolfenstein parameters obtained from a global fit
to the available (mainly $B$-physics) data \cite{PDG},
$\lambda = 0.22506$, $A = 0.811$, $\rho = 0.12714$ and $\eta = 0.365016$,
yielding $\gamma= 70.8^{\scriptstyle o}$, so that  
$r_b\,\sin\gamma \simeq 0.633\times 10^{-3}$\,.

Before using our results it is instructive to estimate the 
combination of hadronic parameters entering Eq.~(\ref{eq:deltaAcp}).
It can readily be extracted from the most accurate LHCb result for 
$\Delta a_{CP}^{dir}$ presented in Eq.~(\ref{eq:acpdel_LHCb}). 
By substituting $r_b\,\sin\gamma$ into 
Eq.~(\ref{eq:deltaAcp}) and neglecting the corrections of $O(r_b^2)$, we obtain 
the following interval for the penguin contribution parameters,
\begin{eqnarray}
0.12 \leq \left(r_K\, \sin\delta_K + r_\pi\, \sin\delta_\pi\right) \leq 1.46\,,
\label{eq:rpiKexp}
\end{eqnarray}
where we allowed for one standard deviation for the experimental data
adding in quadrature the errors quoted in Eq.~(\ref{eq:acpdel_LHCb}).
We see that the combination of amplitudes extracted from experiment 
is quite uncertain. Still, within $1\sigma$-deviation, quite large values of 
the relative magnitudes of penguin effects are allowed, especially if the 
phases are small.

To predict the ratios $r_K$ and $r_\pi$ from our calculations 
we extract the amplitudes $\left|{\cal A}_{\pi\pi}\right|$ and $\left|{\cal A}_{KK}\right|$ 
relating them via Eq.~(\ref{eq:decomp1}) (in which we neglect the small $O(\lambda_b)$ 
terms on the right-hand side) to the decay amplitudes. For determination of the latter, 
we use the experimentally measured branching fractions \cite{PDG},
\begin{eqnarray}
{\cal B}(D^0 \to \pi^-\pi^+) = (1.407\pm 0.025)\times 10^{-3}\,, ~~
{\cal B}(D^0 \to K^-K^+) = (3.97 \pm 0.07)\times 10^{-3}\, .
\label{eq:BRexp}
\end{eqnarray}
Inverting Eq.~(\ref{eq:width}), and using the above values together with the lifetime 
$\tau_{D0}=0.4101$ ps, we obtain
\begin{eqnarray}
 \left|{\cal A}_{\pi\pi}\right| &\simeq& \lambda_s^{-1} \left|A(D\to \pi^-\pi^+)\right| \ \ 
= (2.10\pm 0.02 )\times10^{-6} ~\mbox{GeV}\,, 
\nonumber\\
\left|{\cal A}_{KK}\right| &\simeq& \lambda_s^{-1} \left|A(D\to K^-K^+)\right| 
= (3.80\pm 0.03)\times10^{-6}~\mbox{GeV} \,.
\label{eq:Ampl}
\end{eqnarray}
Finally, we use the estimated penguin hadronic matrix elements in Eqs.~(\ref{eq:rPres_pi})
and (\ref{eq:rPres_K}) to predict the ratios: 
\begin{eqnarray}
r_\pi = \frac{|{\cal P}_{\pi\pi}^{s}|}{|{\cal A}_{\pi\pi}|}= 0.093 \pm 0.011\,, \qquad
r_K = \frac{|{\cal P}_{KK}^{d}|}{|{\cal A}_{KK}|}= 0.075 \pm 0.015\,.
\label{eq:RatAmpl}
\end{eqnarray}
Note that predicting the relative phase $\delta_{\pi}$ or  $\delta_K$
of the total amplitude vs. penguin contribution is a difficult task which is beyond the 
calculation performed here. Although we obtained a certain prediction for the phase 
of the penguin contribution, the phase of the main amplitude still remains obscure. 
One substantial complication is a possible influence of nearby light-quark 
scalar resonances on the strong phases (see, e.g. \cite{Falk:1999ts}). 
Those resonances, however, are known to be rather broad, overlapping in the energy 
region of the $D$-meson mass. This gives us confidence that quark-hadron duality ansatz 
provides a reasonable approximation to the final result.

Substituting our estimates for $r_\pi$ and $r_K$  in 
Eqs.~(\ref{eq:acp1}), (\ref{eq:acp2}) and (\ref{eq:deltaAcp}) taken to $O(r_b)$ 
and allowing for arbitrary strong phases $\delta_{\pi,K}$, we obtain the upper bounds,
\begin{eqnarray} \label{eq:AcpNum1}
\left |a_{CP}^{dir}(\pi^-\pi^+)\right | < 0.012\pm 0.001\%,~~ 
\left |a_{CP}^{dir}(K^-K^+) \right| < 0.009\pm 0.002 \%.
\nonumber \\
\left|\Delta a_{CP}^{dir}\right|  < 0.020\pm 0.003\%\,. ~~~~~~~~~~~~~~
\end{eqnarray}
Alternatively, assuming that the phases $\delta_\pi$ and $\delta_K$ are given 
by the phases of ${\cal P}_{\pi\pi}^{s}$ and ${\cal P}_{KK}^{d}$ calculated above
and presented in Eq.~(\ref{eq:hme_num}), we obtain
\begin{eqnarray}\label{eq:AcpNum2}
a_{CP}^{dir}(\pi^-\pi^+) &=& -0.011\pm 0.001\%, 
\nonumber \\
a_{CP}^{dir}(K^-K^+) &=& \phantom{-}0.009 \pm 0.002\%.
\\
\Delta a_{CP}^{dir}&=&\phantom{-}0.020 \pm 0.003\% \,.
\nonumber
\end{eqnarray}
This is equivalent to assuming that the dominant parts of the decay amplitudes 
parametrized according to Eq.~(\ref{eq:decomp1}) as 
$\lambda_s{\cal A}_{\pi\pi}$ and $\lambda_s{\cal A}_{KK}$ 
both have a small phase relative, respectively, 
to  ${\cal P}_{\pi\pi}^{s}$ and ${\cal P}_{KK}^{d}$.
Such a situation is realized, for example, in a very simplified scenario 
when the decay amplitudes are dominated by the factorization ansatz.
Yet,  this might not be a very reliable approximation, 
as the decompositions (\ref{eq:calApiK}) of 
${\cal A}_{\pi\pi}$ and ${\cal A}_{KK}$
contain  hadronic matrix elements with different topologies, including 
the penguin-topology ones.

Our predictions (\ref{eq:AcpNum1}) and 
(\ref{eq:AcpNum2}) are consistent with the experimental 
results quoted in Eqs.~(\ref{eq:acpdel_PDG}), (\ref{eq:acpdel_LHCb}), and
(\ref{eq:acppi_LHCb}). Note however, that the predicted  
upper bound on the SM contribution to $\Delta a_{CP}^{dir}$ is about 
factor of five smaller than the  magnitude of the central value of 
the currently available experimental interval (\ref{eq:acpdel_LHCb}).

\section{Conclusion}\label{conclusions}

In this letter, we presented a new method to estimate the key hadronic matrix elements 
determining the direct CP asymmetries and their difference in $D^0\to  K^- K^+
$ and $D^0\to\pi^-\pi^+$ nonleptonic decays. The method is a variant of the 
QCD-based LCSR technique adopted from the computations of $B\to \pi\pi$ decay 
amplitudes. A nontrivial strong rescattering phase emerges in 
the calculated hadronic matrix elements. Our results 
do not rely on any flavor-symmetry and/or model-inspired 
amplitude decomposition. 
They do, however, rely heavily on the assumption of 
quark-hadron duality, which introduces a yet unaccounted systematic uncertainty.

An interesting question, directly related to the duality violation 
is the influence of intermediate 
scalar-isoscalar $f_0$ resonances on the decay amplitudes.
One possibility to address this question 
is to modify our method in the following way. Instead of directly 
continuing the calculated 
correlation function to the physical point $P^2=m_D^2$, one can match 
the result  to the hadronic dispersion relation in the variable 
$P^2$ at spacelike $P^2<0$, adopting a certain pattern of 
resonances in this dispersion relation. The hadronic matrix element is then 
obtained by setting $P^2=m_D^2$ in the fitted dispersion relation.  
This, more involved version of the LCSR method is postponed to a future study. 

Our main results are the ratios (\ref{eq:RatAmpl}) of the  calculated ``penguin topology'' 
matrix elements to 
the total $D^0\to P^- P^+$ decay amplitudes, extracting the absolute values 
of the latter amplitudes
from experimental data on $D^0$-decay rates to charged pions and kaons. 

The upper bounds (\ref{eq:AcpNum1}) and estimates (\ref{eq:AcpNum2})
obtained here for the direct CP-asymmetry  in both pion and kaon modes 
and their difference
quantitatively assess the expected amount of direct CP violation in 
the charm sector of Standard Model. We believe that our results 
will become useful for the interpretation of current and future measurements
of this elusive effect.\\

{\bf Acknowledgements}

The work of A.K. is supported by DFG 
Research Unit FOR 1873 ``Quark Flavour Physics and Effective Field Theories", 
contract No KH205/2-2.
A.A.P. is supported in part by the U.S. Department of Energy under contract DE-SC0007983. 
A.A.P. is a Comenius Guest Professor at the University of Siegen.


\end{document}